\documentclass{sig-alternate}

\usepackage[utf8]{inputenc}
\usepackage{amsmath}
\usepackage{amsfonts}
\usepackage{mathtools}

\usepackage{xcolor}
\usepackage{booktabs}

\def \calG {\mathcal{G}}
\def \calN {\mathcal{N}}
\def \calE {\mathcal{E}}
\def \calL {\mathcal{L}}

\newcommand{\mtrx}[1]{{#1}}
\newcommand{\vect}[1]{{#1}}

\makeatletter
\def\@copyrightspace{\relax}
\makeatother

\begin{document}

\title{Inference of Users Demographic Attributes based on Homophily in Communication Networks}

\numberofauthors{3}

\author{
\alignauthor
Jorge Brea\\
       \affaddr{Grandata Labs}\\
       \affaddr{Buenos Aires, Argentina}\\
       \texttt{jorge@grandata.com}
\alignauthor
Javier Burroni\\
       \affaddr{Grandata Labs}\\
       \affaddr{Buenos Aires, Argentina}\\
       \texttt{javier.burroni@grandata.com}
\alignauthor
Carlos Sarraute\\
       \affaddr{Grandata Labs}\\
       \affaddr{Buenos Aires, Argentina}\\
       \texttt{charles@grandata.com}
}

\date{}

\maketitle

\section{Introduction}

Over the past decade, mobile phones have become prevalent in all
parts of the world, across all demographic backgrounds. Mobile phones
are used by men and women across a wide age range in both
developed and developing countries.
Consequently, they have become one of the most important mechanisms for
social interaction within a population, making them an increasingly
important source of information to understand human demographics and
human behaviour. 

In this work we combine two sources of information: 
communication logs from a major mobile operator in a Latin American country,
and information on the demographics of a subset of the users
population.
This allows us to perform an observational study of 
mobile phone usage, differentiated by age groups categories \cite{brea2014,sarraute2014}.
This study is interesting in its own right, 
since it provides knowledge on the structure and
demographics of the mobile phone market in the studied country.

We then tackle the problem of inferring the age group for all users in the network. 
We present here an exclusively graph-based inference method 
relying solely  on the
topological structure of the mobile network, together with a topological analysis of the 
performance of the algorithm.
The equations for our algorithm can be described as a diffusion process with two added properties: 
(i) memory of its initial state, and (ii) the information is propagated as a probability vector 
for each node attribute (instead of the value of the attribute itself).  
Our algorithm can successfully infer different
age groups within the network population given known values for a subset of nodes
(seed nodes). Most interestingly, we show
that by carefully analysing the topological relationships between correctly 
predicted nodes and the seed nodes, we can characterize particular subsets
of nodes for which our inference method has significantly higher accuracy.

\section{Data Set}

The dataset used in this work consists of anonymized call detail records (CDR) and SMS (short message
service) records collected over a three month period.
The dataset contains no personal information.
We aggregate this information into an edge list $(x,y,w_{x,y})$
where $w_{x,y}$ is a boolean value
indicating whether users $x$ and $y$ have communicated at least once within the 
three month period. This edge list represent our mobile phone graph  
$\calG = \left< \calN, \calE \right> $ where $\calN$ denotes the set of nodes (users) 
and $\calE$ the set of communication links. 
Our graph has about 70 million nodes and 250 million edges.
We are also given the age of a subset of 500,000 nodes, 
which we use as a ground truth (denoted $\calN_{GT}$).

\section{Age Homophily in the Communication Network}

\begin{figure}[ht]
\centering
	{\includegraphics[trim=1.5cm 0cm 1.5cm 1.0cm, clip=true, width=0.95\linewidth]
	{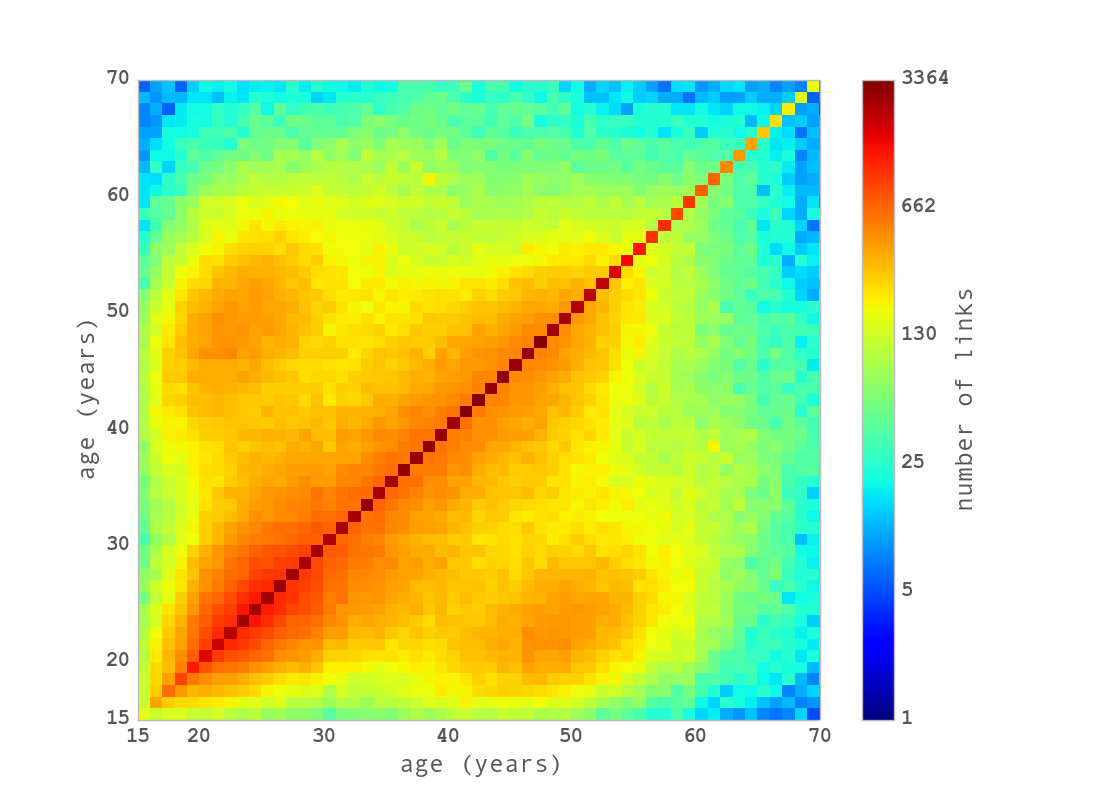}}
	(a) Communications matrix $C$.
	\centering
	{\includegraphics[trim=1.5cm 0cm 1.5cm 1.0cm, clip=true, width=0.95\linewidth]
	{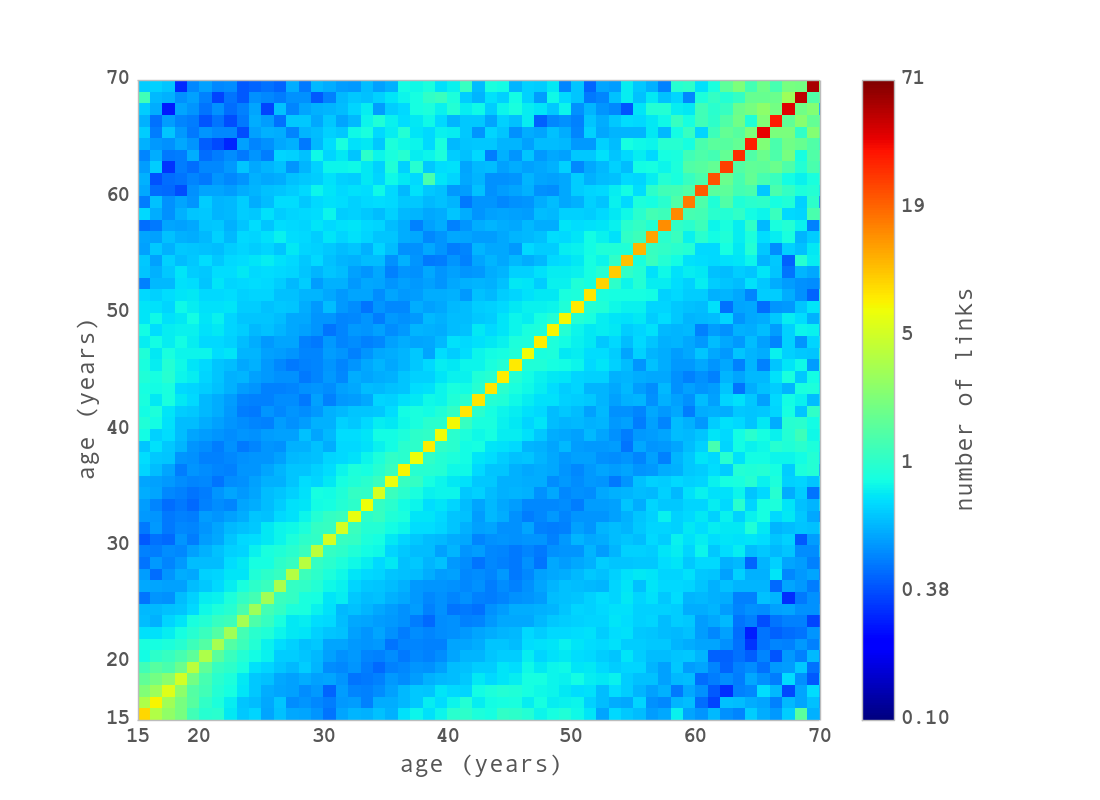}}
	(b) Difference between $C$ and $R$.
	
	\caption{Age homophily plots showing (a) the communications matrix $C$ and 
	(b) the difference between $C$ and the surrogate random links matrix $R$.}
	\label{fig:communications}
\end{figure}

Graph based methods like the one we present in this work rely strongly on the ability of the 
graph topology to capture correlations between the node attributes we are aiming 
to predict. A most fundamental structure is that of correlations between a node's attribute and that of its neighbors. Figure~\ref{fig:communications}(a) shows the correlation matrix $\mtrx{C}$ 
where $\mtrx{C_{i,j}}$ is 
the number of links between users of age $i$ and age $j$ for the nodes in the ground
truth $\calN_{GT}$. Though we can observe some smaller off diagonal peaks, we can see that it is 
most strongly peaked along the diagonal,
showing that users are much more likely to communicate with users of their same age. 

To account for a population density bias, 
we compute a surrogate correlation matrix $\mtrx{R}$ 
as the expected number of edges
between ages $i$ and $j$ under the null hypothesis of independence.
This matrix represents a graph with the same nodes as the original, but with random edges
(while maintaining the same number of edges as the original). 
Both $\mtrx{C}$ and $\mtrx{R}$ are represented with a logarithmic color scale.
If we subtract the logarithm of $\mtrx{R}$ to the logarithm of the original matrix $\mtrx{C}$, 
we can isolate the ``social effect'' (i.e. homophily) from the pure random connections, 
as can be seen in Fig.~\ref{fig:communications}(b).

\section{Reaction-Diffusion Algorithm}

For each node $x$ in $\calG$ we define an initial state probability vector $\vect{g}_{x,0} \in \mathbb{R}^{4}$
representing the initial probability of the nodes age belonging to one of $d=4$ age categories: 
below 24, 25 to 34, 35 to 50, 
and over 50 years old. 
More precisely, each component of $\vect{g}_{x,0}$ is given by
\begin{equation}
\label{eq:inititalconditions}
	(\vect{g}_{x,0})_i = \begin{cases}
		\delta_{i,a(x)} & \text{if }  x \in \calN_{S}  \\
		1/d  & \text{if } x \not\in \calN_{S} 
	\end{cases}
\end{equation}
where $\delta_{i,a(x)}$ is the Kronecker delta function, 
$a(x)$ the age category assigned to each seed node $x$, 
and $\calN_S \subset \calN_{GT}$ are the seed nodes, whose known age attribute is diffused across the network. For non-seed nodes, equal probabilities are assigned to 
each category. These probability vectors are the set of initial conditions for the algorithm.

The evolution equations for the probability vectors $\vect{g}$ are then as follows:
\begin{equation}
\label{eq:difusiongeneral_con_pesos}
	\begin{split}
		\vect{g}_{x,t} = (1-\lambda) \; \vect{g}_{x,0}+ \lambda \; \frac{\sum_{x\sim y} \vect{g}_{y,t-1} }{|\{y:x\sim y\}|} 
	\end{split}
\end{equation} 
where $x \sim y$ is the set of $x$'s neighbours and $\lambda \in [0,1]$ defines the relative importance of each of the two terms. 
It is not hard to show \cite{brea2014} that the above equation can be rewritten as a discrete reaction diffusion equation on the mobile phone graph given by
\begin{equation}\label{eq:laplacian}
	\vect{g}_{t}^{a}-\vect{g}_{t-1}^{a} = (1-\lambda)(\vect{g}_{0}^{a}-\vect{g}_{t-1}^{a}) 
	- \lambda \mtrx{\calL} \vect{g}_{t-1}^{a}
\end{equation}
where $\calL$ is a normalized graph Laplacian. At each iteration, each
$\vect{g}_{x,t}$ in~\eqref{eq:difusiongeneral_con_pesos} updates its state
as a result of its initial state and the mean field resulting from the probability
vector of its neighbors.

\subsection*{Preserving Age Demographics}

A salient feature of our algorithm is that the demographic information being spread 
is not the age group itself, but a probability vector for each age group.
In each iteration, the algorithm does not collapse the information in each node
to a preferred value; instead it allows the system to evolve as a probability state over the network, which allows us to impose further external constraints on the algorithm's results. For instance, after the last iteration, we can select the age category of each node based on the final probability state of each node's category, but constrained so that the age group distribution for the whole network be that of the ground truth set $\calN_{GT}$ as was described in \cite{brea2014,sarraute2014}.

\section{Results}

In this section we first present the results for the predictive power
of the reaction-diffusion algorithm over the whole network $\calG$. The overall performance for the
entire validation set ($20\%$ of $\calN_{GT}$) was $46.6\%$, and we note that a performance based on random
guessing without prior information would result in an expected performance of $25\%$,
or an expected performance of $\sim 36\%$ if we set all nodes age group to the most 
probable category ($35-50$). We now take a closer look and see how the performance can increase or decrease 
as we select particular subsets of nodes.

\subsection*{Topological Metrics and Performance}

We first look at how our algorithm performs for
different subsets of $\calN$ selected according to  
three topological metrics:
(i) number of seeds in the node's neighborhood, 
(ii) distance of node to $\calN_S$
and (iii) node's degree.

\begin{figure}[th]
    \centering
    {\includegraphics[trim=1.5cm 0cm 1.5cm 1.0cm, clip=true, width=0.95\linewidth]
    {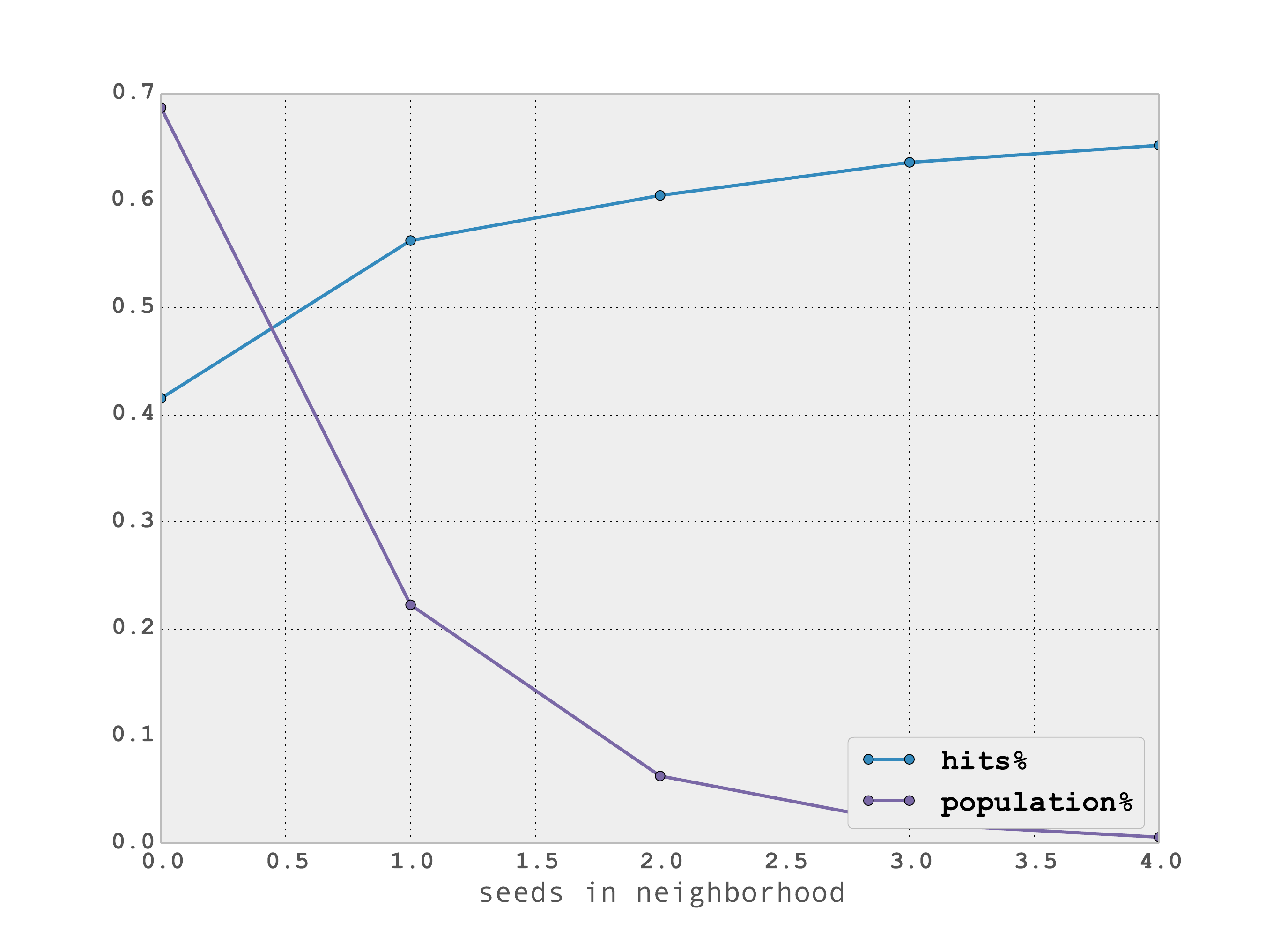}}
    \caption{Performance and population as function of \emph{SIN} (seeds in neighborhood).}
    \label{fig:seedsinego}
\end{figure}

Figure~\ref{fig:seedsinego} plots the performance (\emph{hits}) of
the algorithm 
as a function of 
the number of seeds in the node's neighborhood (\emph{SIN}). 
The algorithm performs worse for nodes with no seeds in the immediate neighborhood with \emph{hits} = 41.5\%,
steadily rising as the amount of seeds increase
with a performance of \emph{hits} = 66\% for nodes with 4 seeds in their neighbourhood.
We also see that the population of nodes decreases exponentially
with the amount of seeds in their neighbourhood. 

\begin{figure}[t]
    \centering
    {\includegraphics[trim=1.5cm 0cm 1.5cm 1.0cm, clip=true, width=0.95\linewidth]
    {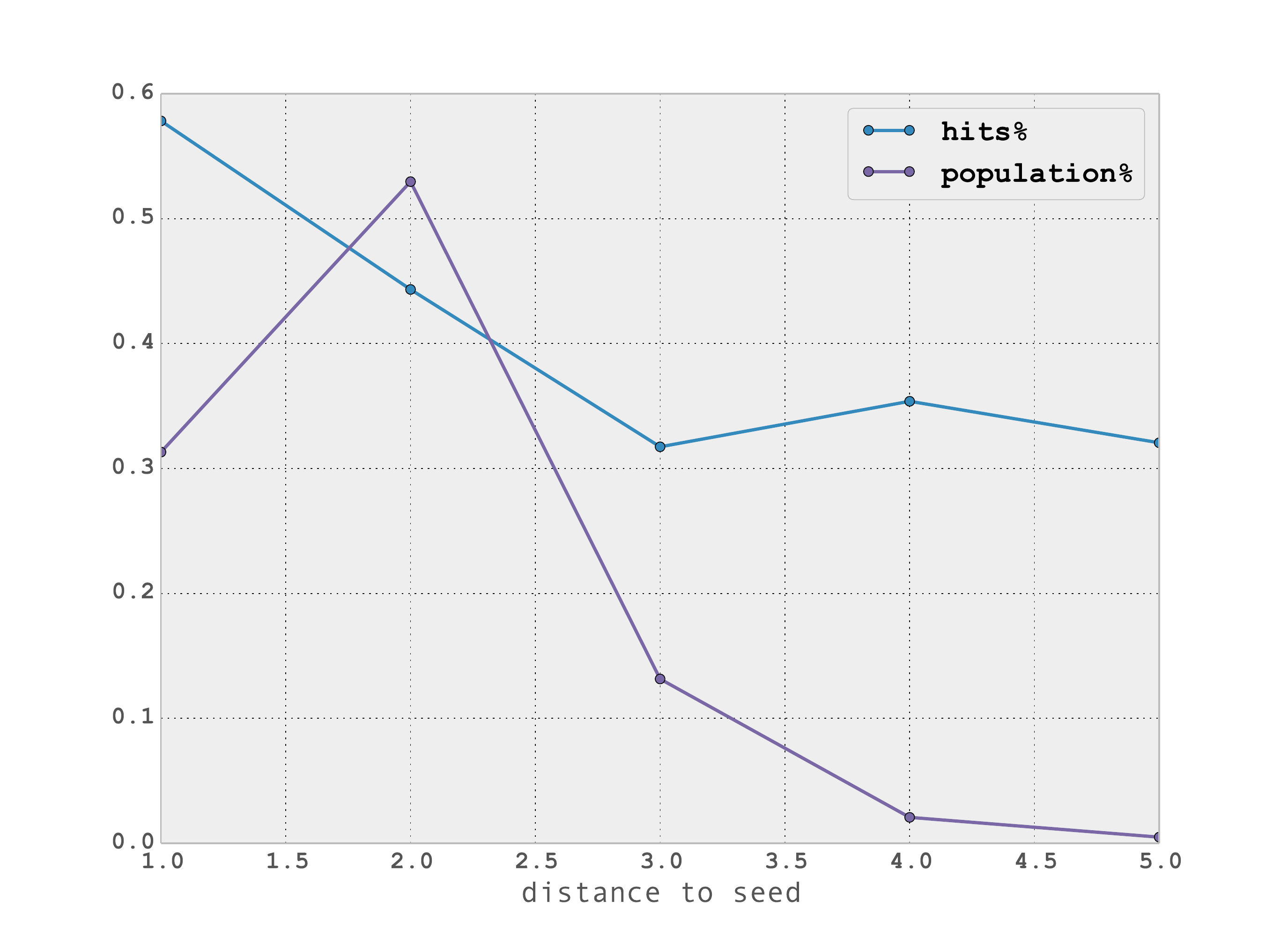}}
    \caption{Performance and population as function of \emph{DTS} (distance to seeds set).}
    \label{fig:distancetoseeds}
\end{figure}

Next we examine how the algorithm performs
for nodes in $\calG$ that are at a given distance to the seed set (\emph{DTS}).
In Fig.~\ref{fig:distancetoseeds} we plot the population size
of nodes as a function of their \emph{DTS}. The most frequent
distance to the nearest seed is 2, and almost all nodes
are at distance less than 4. This implies
that after four iterations of the algorithm, the seeds information
have spread to most of the nodes in $\calG$. This figure also shows
that the performance of the algorithm decreases as the distance of a node
to $\calN_S$ increases.

\begin{figure}[t]
	\centering
    {\includegraphics[trim=1.7cm 0cm 2.0cm 1.0cm, clip=true, width=0.95\linewidth]
	{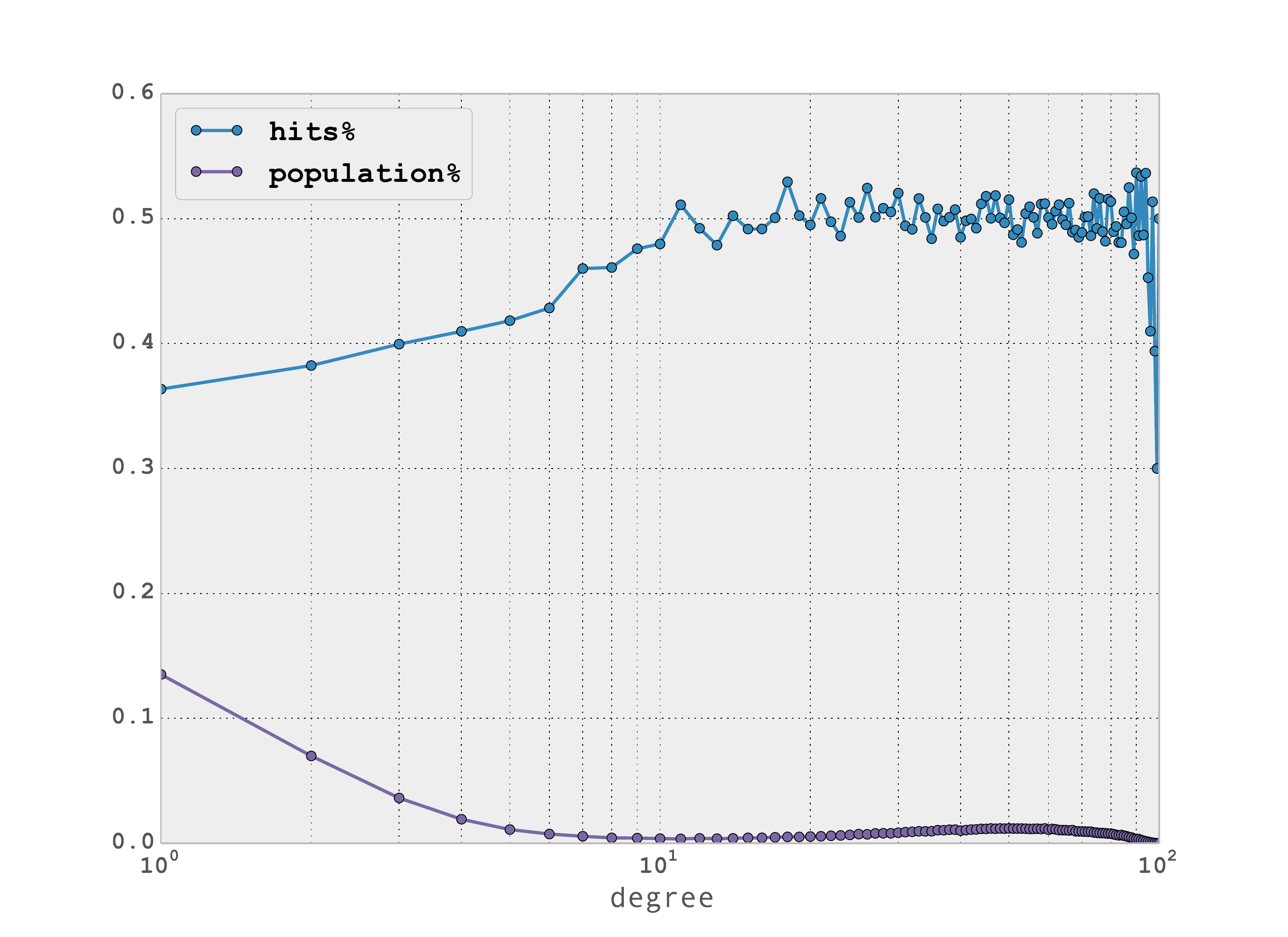}}
	\caption{Performance and node population as function of the nodes degree.}
	\label{fig:nodesdegree}
\end{figure}

In Fig.~\ref{fig:nodesdegree} we see that the performance
of the algorithm is lowest for nodes with small degree and gradually increases as
the degree increases reaching a plateau for nodes with $d(x) > 10$. 

\subsection*{Probability Vector Information}

An orthogonal approach to find an optimal subset of 
nodes (where our algorithm works best) is to exploit the information in the probability vector 
for the age group on each node. Namely, we examine the performance 
of the reaction-diffusion algorithm as we restrict our analysis  
to nodes whose selected category satisfies a minimal 
threshold value $\tau$ in its probability vector. 

\begin{figure}[t]
	\centering
    {\includegraphics[trim=1.0cm 0cm 0cm 1.0cm, clip=true, width=0.95\linewidth]
	{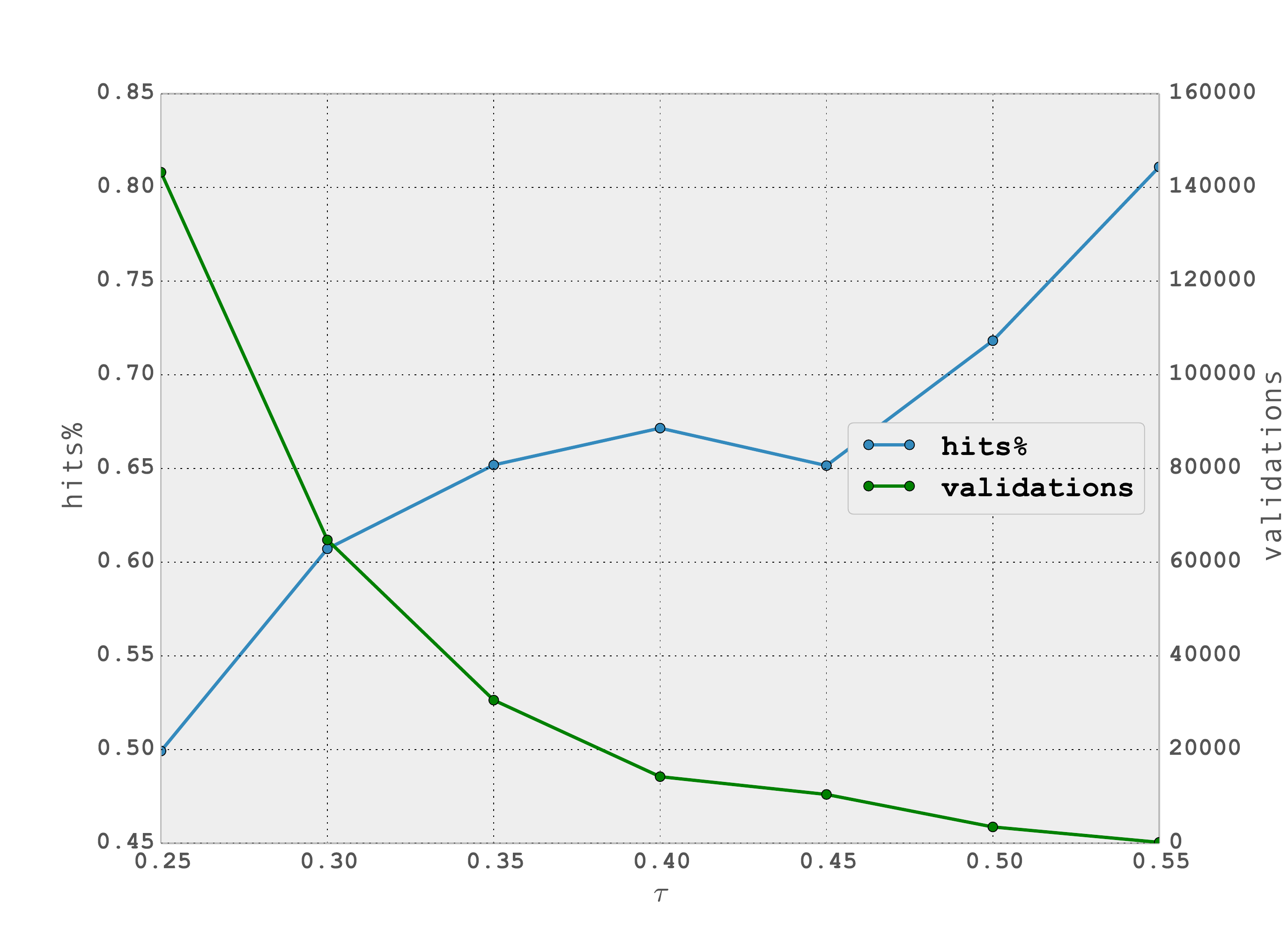}}
	\caption{Performance as function of $\tau$.}
	\label{fig:max_prob_telco}
\end{figure}

In Fig.~\ref{fig:max_prob_telco}
we observe a monotonic increase in the performance as the threshold is increased 
but, as expected there is a monotonic decrease of the validation set size. We note
that for $\tau=0.5$ the performance increased to $72\%$ 
with 3,492 out of the 143,240 (2.4\%) of the validation nodes remaining. 
The performance of the algorithm increases to $\sim 81\%$ for $\tau=0.55$ 
but the validation set remaining sharply decreased to only 201 nodes ($0.1\%$).

\section{Conclusions}

In this work we have presented a novel algorithm that can harness the bare bones topology of mobile phone
networks to infer with significant accuracy the age group of the network's users.
We show that an important reason for the success of the algorithm is the strong age homophily among neighbours in the network as evidenced by our observational study of the ground truth sample $\calG_{GT}$.   

We have shown the importance
of understanding nodes topological properties, in particular their relation to the seed nodes,  
in order to fine grain our expectation of correctly classifying the nodes. 
Though we have carried out this analysis for a specific network, 
we believe this approach can be useful to study generic networks where node attribute correlations are present.

As future work, one direction that we are investigating
is how to improve the graph based inference approach presented here by
appropriately combining  it with classical machine learning techniques based on node features \cite{sarraute2014}.  
We are also interested in applying our methodology 
to predict variables related to the users' spending behavior.
In \cite{singh2013predicting} the authors show correlations between social features
and spending behavior for a small population.
We are currently tackling the problem of predicting spending behavior characteristics
on the scale of millions of individuals.


\bibliographystyle{abbrv}
\bibliography{../bibliography/sna}
\end{document}